\documentclass{article}
\usepackage{dcase2022,amsmath,amsfonts,graphicx,url,times,booktabs, tabularx}
\usepackage{multirow}
\usepackage{float}
\usepackage{xcolor}

\title{
Few-shot bioacoustic event detection \\
at the DCASE 2022 challenge
}

\name{I. Nolasco$^{1}$,
      S. Singh$^{1}$,
      E. Vida\~{n}a-Villa$^{9}$,
      E. Grout$^{3,4}$,
      J. Morford$^{10}$,
      M. Emmerson$^{11}$
       F. Jensens$^{12}$, 
      }
\secondlinename{	
       H. Whitehead$^{13}$,
       I. Kiskin$^{14}$
    A. Strandburg-Peshkin$^{3,4}$,
    L. Gill$^{5}$, 
      H. Pamu\l{}a$^{6}$, 
      V. Lostanlen$^{2}$, 
      V. Morfi$^{15}$,
      D. Stowell$^{8}$
      }
\address{
$^1$ Centre for Digital Music (C4DM), Queen Mary University of London, London, UK\\
$^2$ Nantes Université, École Centrale Nantes, CNRS, LS2N, UMR 6004, F-44000 Nantes, France \\
$^3$ Dept. of Biology \& Centre for the Advanced Study of Collective Behaviour, University of Konstanz, Germany\\
$^4$ Dept. for the Ecology of Animal Societies, Max Planck Institute of Animal Behavior, Germany\\
$^5$ BIOTOPIA Naturkundemuseum Bayern, Munich, Germany\\
$^6$ AGH University of Science and Technology, Kraków, Poland\\
$^7$ Cornell Lab of Ornithology, Cornell University, Ithaca, NY, US\\
$^8$ Tilburg University, Tilburg, The Netherlands; Naturalis Biodiversity Centre, Leiden, The Netherlands\\
$^9$ La Salle, University Ramon Llull, Barcelona, Spain\\
$^{10}$ The Oxford Navigation group, Dept. of Zoology, Oxford University, Oxford, UK \\
$^{11}$ School of Biology and Behavioural Sciences, Queen Mary University of London, London, UK \\
$^{12}$ College of Arts and Sciences at Syracuse University,  Syracuse, NY, US\\
$^{13}$ School of Science, Engineering and Environment, University of Salford, Manchester, UK\\
$^{14}$ Institute for People-Centred AI, FHMS,
University of Surrey, Surrey, UK\\
$^{15}$ Sonantic Limited, London, UK
 }

\begin{document}

\ninept
\maketitle

\begin{sloppy}

\begin{abstract}

Few-shot sound event detection is the task of detecting sound events, despite having only a few labelled examples
of the class of interest. This framework is particularly useful in bioacoustics, where often there is a need to annotate very long recordings but the expert annotator time is limited.
This paper presents an overview of the second edition of the few-shot bioacoustic sound event detection task included in the DCASE 2022 challenge.
A detailed description of the task objectives, dataset, and baselines is presented, together with the main results obtained and characteristics of the submitted systems.
This task received submissions from 15 different teams from which 13 scored higher than the baselines. The highest F-score was of 60\% on the evaluation set, which leads to a huge improvement over last year's edition. Highly-performing methods made use of prototypical networks, transductive learning, and addressed the variable length of events from all target classes. Furthermore, by analysing results on each of the subsets we can identify the main difficulties that the systems face, and conclude that few-show bioacoustic sound event detection remains an open challenge.
%The highest score achieved 60\% F-score on the evaluation set.
%This means a huge improvement over last year's edition, however by analysing results on each of the subsets of the evaluation set it is possible to verify what are the main difficulties that these systems face, and realise that few-shot bioacoustic sound event detection remains an open challenge.

\end{abstract}

\begin{keywords}
Few-shot learning, bioacoustics, sound event detection, DCASE challenge
\end{keywords}

\vspace{-0.2cm}
\section{Introduction}
\label{sec:intro}
\vspace{-0.2cm}
% Bioacoustic SED
The task of bioacoustic sound event detection refers to the retrieval of animal vocalisations from audio recordings in terms of onset and offset times.
It shares a common methodology with other sound event detection (SED) contexts, %such as offices %\cite{stowell2015detection}, homes \cite{turpault2019sound}, city streets \cite{mesaros2010acoustic}, and high-security spaces \cite{mesaros2019sound}.
yet, the application domain of bioacoustics is particularly challenging for SED. % in part because of the high diversity of possible recording conditions and of vocalisation types \cite{stowell2018computational}.
Deep learning contributed to overcome some of these difficulties in bioacoustic SED, however it also established strong requirements regarding the amount of annotated data needed \cite{stowell2022deeplearning}.
Collecting and annotating a large dataset of animal vocalisations is often not feasible given that species are unequally abundant \cite{vellinga2015xeno} and may be rarely observed; and audio annotation is costly and time-consuming \cite{mendez2019machine}.%; and, more fundamentally, the taxonomy may vary depending on the use case \cite{cramer2020chirping}.
% Why Few shot
In contrast to traditional deep learning approaches that use a large amount of data to train models, few-shot learning tries to build accurate models with very few training data \cite{Snell2017PrototypicalNF}.
Few-shot learning is usually studied using $N$-way-$k$-shot classification, where $N$ denotes the number of classes and $k$ the number of known examples for each class.

% task setup and goal
This problem was first evaluated as a task on the DCASE 2021 challenge. This year, the setup and goal remain the same: Given the first 5 events of a target class, can systems detect the subsequent events of the same class in the remaining of the audio recording?
% state of the art approaches to Few-shot
Diverse approaches have been used to address the few-shot learning problem for classification. Some use prior knowledge about similarity between sounds by computing embeddings (learnt representation spaces) designed to help discriminate between unseen classes \cite{Snell2017PrototypicalNF}, while others exploit prior knowledge about the structure of the data by using augmentation to synthesise new data \cite{Wang2018LowShotLF}. Finally, some approaches can learn models with parameters that can be fine-tuned to smaller datasets \cite{Nichol2018OnFM}. 
More recent works use meta-learning and/or prototypical networks for acoustic few-shot learning \cite{Wang2020}, \cite{shiu2020deep}. All of the above approaches deal with classification tasks rather than detection. Indeed, SED in a few shot setup is commonly approximated as an audio tagging task and few works have addressed the actual detection of onsets and offsets of events \cite{wolters2021proposal}. 
% contributions, last versus this year task
At last year's task edition, the best ranked system improved over the baseline prototypical approach by applying a transductive inference method and a mutual learning framework designed to make the feature extraction network more task dependent  \cite{Zou2021}.
The overall best results were just bellow 40\% f-score which indicates the difficulty of this task. This year, we added more and diverse datasets, and increased the task difficulty (dataset diversity); yet the task doubled the amount of participants and the best overall f-score in the evaluation set reached the 60\% level.
% delineate paper
This paper is structured as follows. Section \ref{sec:datasets} presents the bioacoustic datasets used for developing and evaluating submitted systems. Section \ref{sec:baseline} presents the two baseline methods proposed for the task, followed by the evaluation procedure. Finally, section \ref{sec:results} presents the results of the submitted systems and a discussion about the overall task and future steps in the field of few-shot bioacoustic event detection.

% \begin{figure}[t]
%     \centering
%     \includegraphics[width=0.8\columnwidth]{VM.png}
%     \caption{Overview of the proposed few-shot bioacoustic event detection task at the DCASE challenge. Green and purple rectangles represent labelled and predicted events, respectively.}
%     \vspace{-0.5cm}
%     \label{fig:overview}
% \end{figure}
\begin{table*}[h!]
    \centering
    \begin{tabular}{r|c|c|c|c|c|c}
         & Dataset & mic type &  \# audio files & total duration & \# labels (excl. UNK) & \# events \\
         \hline
        \multirow{4}{*}{Development Set: Training} & BV & fixed & 5 & 10 hours & 11 & 9026\\
        & HT & various & 5& 5 hours & 5 & 611\\
        & MT & mobile & 2 & 70 mins & 4 & 1294\\
        & JD & mobile & 1 & 10 mins & 1 & 357\\
        & WMW & various & 161 & 5 hours  & 26 & 2941\\
        \hline
        \multirow{2}{*}{Development Set: Validation} & HB & handheld & 10 & 2.38 hours & 1 & 712 \\
        & PB & fixed &  6 & 3 hours & 2 & 292 \\
        & ME & handheld & 2 & 20 mins & 2 & 73\\
        \hline
        \multirow{3}{*}{Evaluation Set} &  CHE & fixed & 18 & 3 hours & 3 & 2550\\
        & DC & fixed &  10 & 95 mins & 3 & 967\\
        & CT & handheld & 3  & 48 mins  & 3 & 365 \\
        & MS & fixed &  4 & 40 mins & 1 & 1087\\
        & QU & handheld &  8 & 74 mins & 1 & 3441\\
        & MGE & fixed &  3 & 32 mins & 2 & 1195\\
    \end{tabular}
    \caption{Information on each dataset.}
    \label{tab:datasets}
\end{table*}

\vspace{-0.2cm}
\section{Datasets}
\label{sec:datasets}

\vspace{-0.2cm}
A \textit{development dataset} consists of predefined training and validation sets to be used for system development. \footnote{Dev set: \url{https://doi.org/10.5281/zenodo.6012309}}. 
The training set contains  multi-class temporal annotations, provided for each recording as: positive (POS), negative (NEG) and unknown (UNK). For the validation set only single-class temporal annotations (POS/UNK) were provided for each recording. 
A separate \textit{evaluation set} was kept for evaluating the performance of the systems.\footnote{Eval set: \url{https://doi.org/10.5281/zenodo.6517413}} 
During the task, only the first five POS events of the class of interest were provided for each of the recordings. 
Table \ref{tab:datasets} presents an overview of all the datasets in the development and evaluation sets.

\textbf{BirdVox-DCASE-10h (BV):}
%\subsection{BirdVox-DCASE-10h (BV)}
%\label{ssec:birdvox}
The BirdVox-DCASE-10h (BV) contains five audio files from four different autonomous recording units, each lasting two hours.
These autonomous recording units are all located in Tompkins County, NY, US.
They follow the same hardware specification: the Recording and Observing Bird Identification Node (ROBIN) developed by the Cornell Lab of Ornithology \cite{lostanlen2018birdvox}.
All recordings were acquired in 2015, during the fall migration season.
An expert ornithologist, Andrew Farnsworth, has annotated flight calls from four families of passerines, namely: American sparrows, cardinals, thrushes, and New World warblers.
%The annotator found 2,662 flight calls from 11 different species in total.
These flight calls have a duration in the range 50--150 milliseconds and a fundamental frequency in the range 2--10 kHz.

\textbf{Hyenas (HT):}
%\subsection{Hyenas (HT, HV)}
%\label{ssec:hyenas}
Spotted hyenas are a highly social species that live in ``fission-fusion" groups and
%where group members range alone or in smaller subgroups that split and merge over time, using 
use a variety of types of vocalisations to coordinate with one another.
%Hyenas use a variety of types of vocalizations to coordinate with one another over both short and long distances. 
Spotted hyenas were recorded on custom-developed audio tags designed by Mark Johnson and integrated into combined GPS/acoustic collars (Followit Sweden AB) by Frants Jensen and Mark Johnson. Collars were deployed on female hyenas of the Talek West hyena clan at the MSU-Mara Hyena Project (directed by Kay Holekamp) in the Masai Mara, Kenya as part of a multi-species study on communication and collective behavior. Recordings used as part of this task contain a variety of different vocalisations which were identified and classified into types based on the established hyena vocal repertoire \cite{Lehmann2020thesis}. Field work was carried out by Kay Holekamp, Andrew Gersick, Frants Jensen, Ariana Strandburg-Peshkin, and Benson Pion; labeling was done by Kenna Lehmann and colleagues.

\textbf{Meerkats (MT, ME):}
%\subsection{Meerkats (MT, ME)}
%\label{ssec:meerkats}
Meerkats are a highly social mongoose species that live in stable social groups and use a variety of distinct vocalisations to communicate and coordinate with one another. The meerkat vocal repertoire has been well characterized based on previous research, allowing calls to be reliably classified by human labellers \cite{Manser1998thesis, MANSER2014281}. Recordings used in this task were acquired at the Kalahari Meerkat Project (Kuruman River Reserve, South Africa; directed by Marta Manser and Tim Clutton-Brock), as part of a multi-species study on communication and collective behavior. Recordings of the development set (MT) were recorded on small audio devices (TS Market, Edic Mini Tiny+ A77, 8 kHz) integrated into combined GPS/audio collars which were deployed on multiple members of meerkat groups. % to monitor their movements and vocalisations. 
Recordings of the validation set (ME) were recorded by an observer following a focal meerkat with a Sennheiser ME66 directional microphone (44.1 kHz) from a distance of less than 1 m. Recordings were carried out during daytime hours while meerkats were primarily foraging and include several different call types. Field work was carried out by Ariana Strandburg-Peshkin, Baptiste Averly, Vlad Demartsev, Gabriella Gall, Rebecca Schaefer and Marta Manser. Audio recordings were labeled by Baptiste Averly, Vlad Demartsev, Ariana Strandburg-Peshkin, and colleagues.

\textbf{Jackdaws (JD):}
%\subsection{Jackdaws (JD)}
%\label{ssec:jackdaws}
Jackdaws are corvid songbirds that usually breed, forage and sleep in large groups. They produce thousands of vocalisations per day, but many aspects of their vocal behaviour remain unexplored. 
%due to the difficulty in recording and assigning vocalisations to specific individuals.
In a multi-year field study (Max-Planck-Institute for Ornithology, Seewiesen, Germany), wild jackdaws were equipped with small backpacks containing miniature voice recorders (Edic Mini Tiny A31, TS-Market Ltd., Russia) to investigate the vocal behaviour of individuals interacting with their group and behaving freely in their natural environment. %The JD dataset contains a 10-minute on-bird sound recording (22050 Hz) of one male jackdaw during the breeding season 2015. 
Field work was conducted by Lisa Gill, Magdalena Pelayo van Buuren and Magdalena Maier. Sound files were annotated by Lisa Gill.%, based on a previously established video-validation in a captive setting \cite{stowell2017bird}.

\textbf{Western Mediterranean Wetlands Bird Dataset (WMW):} This dataset contains bird sounds from 20 endemic species that are typically inhabitants of the ``Aiguamolls de l'Empordà" natural park in Girona, Spain. Depending on the species, audios contain vocalisations such as bird calls or songs; or sounds such as bill clapping (\textit{Ciconia ciconia} species) or drumming (\textit{Dendrocopos minor} species). The audio files that compose this dataset were originally retrieved from the Xeno-Canto portal% \footnote{\url{ http://www.xeno-canto.org/}}
and were manually cleaned and labelled by Juan Gómez-Gómez, Ester Vidaña-Vila and Xavier Sevillano using the Audacity software \cite{gomez2022}. In Xeno-Canto, audios are labelled by quality from A to E, where A means that the audio quality is excellent and E means that the audio quality is poor. This dataset contains only A and B type files which ensured the exclusion of audios with high background noise.% or audios in which the bird is only heard in the background.  

\textbf{HumBug (HB):} Mosquitoes produce sound both as a by-product of their flight and as a means for communication and mating. Fundamental frequencies vary in the range of 150 to 750 Hz \cite{kiskin2020bioacoustic}. As part of the HumBug project, acoustic data was recorded  with a high specification field microphone (Telinga EM-23) coupled with an Olympus LS-14. The recordings used in this challenge are a subset of the datasets marked as \emph{`OxZoology'} and \emph{`Thailand'} from HumBugDB \cite{kiskin2021humbugdb}\footnote{ \url{https://github.com/HumBug-Mosquito/HumBugDB/}}. 
The recordings contain the sound of
 lab-cultured \emph{Culex quinquefasciatus} mosquitoes from Oxford, UK, and various species captured in the wild in Thailand, placed into plastic cups.%\cite{li2018fast}.%

\textbf{Polish Baltic Sea bird flight calls (PB):}
%\subsection{Polish Baltic Sea bird flight calls (PB)}
%\label{ssec:poland}
The PB dataset consists of bird flight calls record%The recordings are the excerpt from Hanna Pamuła's project, focused on the acoustic monitoring of birds migrating at night along the Polish Baltic Sea coast. 
Three autonomous recording units were used with the same hardware settings (Song Meters SM2, Wildlife Acoustics, Inc). They were deployed close to each other ($<$100m) - near the lake, on the dune, and on the forest clearing - to provide diverse acoustic background. 
%The recordings were acquired during the 2016, 2017 and 2018 fall migration seasons. 
The passerines night flight calls were annotated by Hanna Pamuła.
% The PB dataset is part of the development set used for validation. 
Target classes are belong to: song thrush, \emph{Turdus philomelos} and  blackbird, \emph{Turdus merula}. Event lengths vary between 8 to 400 milliseconds and the usual fundamental frequency range for calls is 5 to 9 kHz.

% \textbf{Macaulay Library (ML):}
% %\subsection{Macaulay Library (ML)}
% %\label{ssec:macaulay}
% The Macaulay Library is a digital archive of images, videos, and sounds from animals.\footnote{Official website: \url{https://www.macaulaylibrary.org/}}
% As of 2021, it contains 175k audio recordings from 10k species of birds and 2k species of amphibians, fish, mammals and insects.
% These recordings are contributed by amateur and professional recordists around the world, and the catalogue is maintained by the Cornell Lab of Ornithology.
% For the DCASE 2021 challenge, one author (DB) curated 17 recordings from the Macaulay Library and annotated them in terms of animal vocalizations.
% Each recording contains calls from a different species: 14 terrestrial mammals (not including hyena or meerkat) and 3 birds (not including passeriformes).
% The average duration of each recording is of the order of one minute and the number of calls per minute varies in the range 10--150.

\textbf{Transfer-Exposure-Effects project (CHE):}
This dataset comes from the Transfer Exposure-Effects (TREE) research project\footnote{\url{https://tree.ceh.ac.uk/}}, % \cite{TREE}
which was funded by the Natural Environment Research Council (NERC), Environment Agency and Radioactive Waste Management Ltd. Data were collected using unattended acoustic recorders (Songmeter 3) in the Chornobyl Exclusion Zone (CEZ) to capture the Chornobyl soundscape and investigate the longterm effects of the Chornobyl accident on the local ecology. To date, the study has captured approximately 10,000 hours of audio from the CEZ. The fieldwork was designed and undertaken by Mike Wood (University of Salford), Nick Beresford (UK Centre for Ecology \& Hydrology) and Sergey Gashchak (Chornobyl Center). Common Chiffchaff (Phylloscopus collybita) and Common Cuckoo (Cuculus canorus) vocalisations were manually annotated and labelled from these recordings by Helen Whitehead (University of Salford)% using Raven Pro 1.6.

\textbf{BIOTOPIA Dawn Chorus (DC):}
%\subsection{BIOTOPIA Dawn Chorus (DC)}
%\label{ssec:dawnchorus}
%Acoustic species recognition is a growing field of interest in ornithological and biodiversity research, because it allows unbiased species detection in the absence of expert observers. 
% Many bird species produce vocalisations during the entire day, but their vocally most active period by far usually occurs around dawn. This natural phenomenon is called \textit{dawn chorus}.
%has received a lot of attention in biological studies.
%and also appears to be the perfect time window for species detection, as it provides the largest likelihood of most individuals of the same and of different species signalling. Yet the sheer complexity of undirected dawn chorus recordings have made automatic species classification extremely difficult, leaving this potentially rich source of acoustically determined species data largely untapped. 
The Dawn Chorus project is a worldwide citizen science and arts project bringing together amateurs and experts to experience and record the dawn chorus at their doorstep.% to draw a global picture of bird biodiversity through sound. 
The DC dataset stems from dawn chorus recordings, made using Zoom H2 recorders at 44100 Hz, at three different locations in Southern Germany (Haspelmoor, Munich’s Nymphenburg Schlosspark, and Nantesbuch), by Moritz Hertel and Rudi Schleich. The vocalisations of three target species were annotated by Lisa Gill (Common cuckoo, \emph{Cuculus canorus}; European robin, \emph{Erithacus rubecula}; Eurasian wren, \emph{Troglodytes troglodytes}).
A challenging aspect of this data is related to  recordings being very busy with various other birds vocalising at the same time.

\textbf{Coati (CT):}
Coatis are small carnivorous mammals where females and young males live in groups of up to 25 individuals. They communicate using a series of different calls however their functions are still relatively unknown. The target calls used in this dataset are growls, chitters and chirp-grunts. Several other call types that might be confused with the targets were captured in the recordings which configures the main challenging aspect of this data.

\textbf{Manx Shearwater (MS):}
% Manx Shearwater (Puffinus puffinus) are medium sized migratory seabirds that live in large colonies typically out of the coast of small islands. They nest in burrows to which they return season after season to breed. The data gathered here is from recordings obtained from 4 different burrows using fixed recorder devices. 
% The target class across all recordings is chick begging calls which typically occur in bouts.
% The recordings can be very noisy at moments due to the presence of parents vocalising inside the burrow and from outside of the burrow noise as well. Additionally these calls are of short duration which configures a certain degree of difficulty to this set of data.
% Recordings were labelled by Joe Morford.
Manx shearwaters are procellariiform  seabirds that breed in dense island colonies in the North Atlantic and winter in the South Atlantic off the South American coast. 
Adult Manx shearwaters make loud, distinctive vocalisations while present at their breeding colony in various contexts: to their partner in their nesting burrow, broadcasting from their burrow, and in flight. %They are socially monogamous birds and a pair can produce a single offspring each year. After the egg hatches, the parents regularly visit the chick to feed them through regurgitation and the chick will beg for food from its parents. 
The target class is Chick begging vocalisations which typically comprise bouts of short, high-pitched ‘peeps’. In a multi-year study, Audiomoth recorders were placed in burrows to record the vocalisations
%of both adult Manx shearwaters and chicks 
during the breeding season. Fieldwork on Skomer Island was undertaken by various members of the Oxford Navigation Group (OxNav) and annotation was carried out by Joe Morford.

\textbf{Dolphin Quacks (QU):}
This data consists in underwater recordings of dolphins in their natural habitat. All files were annotated for the Quack target class which represent calls predominantly used in a social context.
Quack events can be very short (from 30 milliseconds) and vary considerably in the course of a single recording. This is a potential problem when the 5 initial POS events of the target class are not good representations of the class overall. 

\textbf{Chick calls (MGE):}
Chicks of domestic chicken were recorded in the Prepared Minds Lab from Queen Mary university of London\footnote{\url{https://www.preparedmindslab.org/home}}.
The 2 target classes are pleasure and contact calls which the chicks will produce as a response to different stimuli. The data gathered here presents uneven time distribution, and calls are typically short. Additionally, calls present low 
stereotypy which contributes to the challenge associated with this dataset.  
\vspace{-0.1cm}
\section{Baseline Methods and Evaluation}
\label{sec:baseline}
\vspace{-0.2cm}
The baseline systems proposed did not change considerably from last year's edition \cite{morfi2021few}.
Template Matching is based on spectrogram cross-correlation and still commonly used in bioacoustics. This approach scored surprisingly well on last edition evaluation set and thus it remains relevant as a baseline for this task.
The second system proposed is based on prototypical networks which remain the state of the art for few-shot learning \cite{Snell2017PrototypicalNF}. The changes from last year's system are the use of a ResNet, and adapting segment size depending on the target class in the query set. These changes mainly address the problem of high variation of event lengths and create a more adaptive system.
\begin{table*}[ht]
    \centering
    \begin{tabular}{l|l|l|l}
    %Rank &
    Team name & \multicolumn{1}{l|}{\begin{tabular}[c]{@{}l@{}}Evaluation set: \\ F-score \% (95\% CI)\end{tabular}} 
    & \multicolumn{1}{l|}{\begin{tabular}[c]{@{}l@{}} Validation set \\ F-score \% \end{tabular}} &
     \multicolumn{1}{l}{\begin{tabular}[c]{@{}l@{}} Main characteristics \end{tabular}} \\
    \hline
    Du\_NERCSLIP\_2 \cite{Du2022} & 60.22 (59.66-60.70)  & 74.4  & CNN+ProtoNet; Frame-level embeddings; PCEN;    \\
    Liu\_Surrey\_2\cite{Liu2022a} & 48.52 (48.18-48.85)  & 50.03  & CNN+ProtoNet; extra data; PCEN+$\triangle {MFCC}$; several post-process. \\
    Martinsson\_RISE\_1 \cite{Martinsson2022} & 47.97 (47.48-48.40)  & 60  & ResNet+ProtoNet; Ensemble (15) based input size; logMel+PCEN \\
    Hertkorn\_ZF\_2 \cite{Hertkorn2022} & 44.98 (44.44-45.42)  & 61.76  &  CNN; Frequency resolution preserving pooling; various post-process \\
    Liu\_BIT-SRCB\_4 \cite{Liu2022}  & 44.26 (43.85-44.62)  & 64.77 & CNN+ProtoNet; Transductive inference   \\
    Wu\_SHNU\_1 \cite{Wu2022} & 40.93 (40.48-41.30)  & 53.88  &   ResNet+ProtoNet; Continual-learning; spectrogram \\
    Zgorzynski\_SRPOL\_4 \cite{Zgorzynski2022}  & 33.24 (32.69-33.69)  &  57.2 & CNN+Siamese Networks; Emsemble (3) average event-length;  \\
    Mariajohn\_DSPC\_1 \cite{Aaquila2022} & 25.66 (25.40-25.91)  & 43.89 & CNN+ProtoNet; logMel; augmentation with time-shifting and mirroring  \\
    Wilbo\_RISE\_4 \cite{Willbo2022} & 21.67 (21.32-21.97) & 47.94 & ResNet+ProtoNEt; Semi-supervised; Melspect+PCEN; several post-process \\
    Zou\_PKU\_1 \cite{Yang2022} & 19.20 (18.88-19.51)  & 51.99  & CNN+protoNet; mutual information loss; time frequency masking + mixup \\
    Huang\_SCUT\_1 \cite{Huang2022}& 18.29 (18.01-18.56)  & 54.63 & transductive inference + Adapted central difference convolution  \\
    Tan\_WHU\_4 \cite{Tan2022} & 17.22 (16.82-17.55)  & 54.53  & CNN+ProtoNet pretrained; transductive inference; task adaptive features  \\
    Li\_QMUL\_1 \cite{Li2022}&  15.49 (15.16-15.77) & 47.88   & CNN+protoNet; PCEN; time, frequency masking + time warping \\
    baseline-TempMatch \cite{morfi2021few} & 12.35 (11.52-12.75)  & 3.37 &  Spectrogram Cross correlation\\
    baseline-ProtoNet \cite{morfi2021few} & 5.3 (5.1-5.2)  &  28.45 & ResNet+ProtoNet  \\
    Zhang\_CQU\_4 \cite{Zhang2022}& 4.34 (3.74-4.56)  & 44.17  &  CNN+protoNet; Fine tunning with MIMI; PCEN\\
    Kang\_ET\_2 \cite{Kang2022} & 2.82 (2.76-2.87) & - & CNN+ProtoNEt; pretrained ECAPA-TDNN; Fine-tuning; Specaugment  \\
    \end{tabular}
    \caption{F-score results per team (best scoring system) on evaluation and validation sets, and summary of system characteristics. Systems are ordered by higher scoring rank on the evaluation set. }
    \label{tab:teams}
\end{table*} 

% \subsection{Evaluation}
% \label{sec:eval}
% %\subsection{Evaluation method}
The evaluation of this task is based on an event-level F-measure with macro-averaged metric across all classes\cite{morfi2021few}.
A positive match between predicted events and reference is found by applying the Intersection over Union (IoU) with 30\% minimum overlap, followed by Hopcroft-Karp-Karzanov algortithm for  bipartite graph matching.
True positives(TP), False positives(FP) and False negatives(FN) can be computed after the matching step. These are defined as: TP - predicted events that match ground truth events; FP - predicted events that do not match any ground truth events; FN - ground truth events that are not predicted.
Matches to UNK events are ignored from these counts as to not negatively impact the systems that predict these events.
Finally, the F-score metric is computed per dataset in the evaluation set and the harmonic mean over all is reported.

%\begin{figure}[h!]
%    \centering
%    \includegraphics[width=\linewidth, height =5cm]{alternative_view_results_team_subset_evaluation.png}
%    \caption{F-Score results by dataset. Systems are ordered by highest scoring rank on the evaluation set. }
%    \label{fig:Fscore_dataset} 
%\end{figure}

\begin{figure}
    \centering
    \includegraphics[width=\linewidth, trim = 3.2cm 3cm 1cm 0cm, clip]{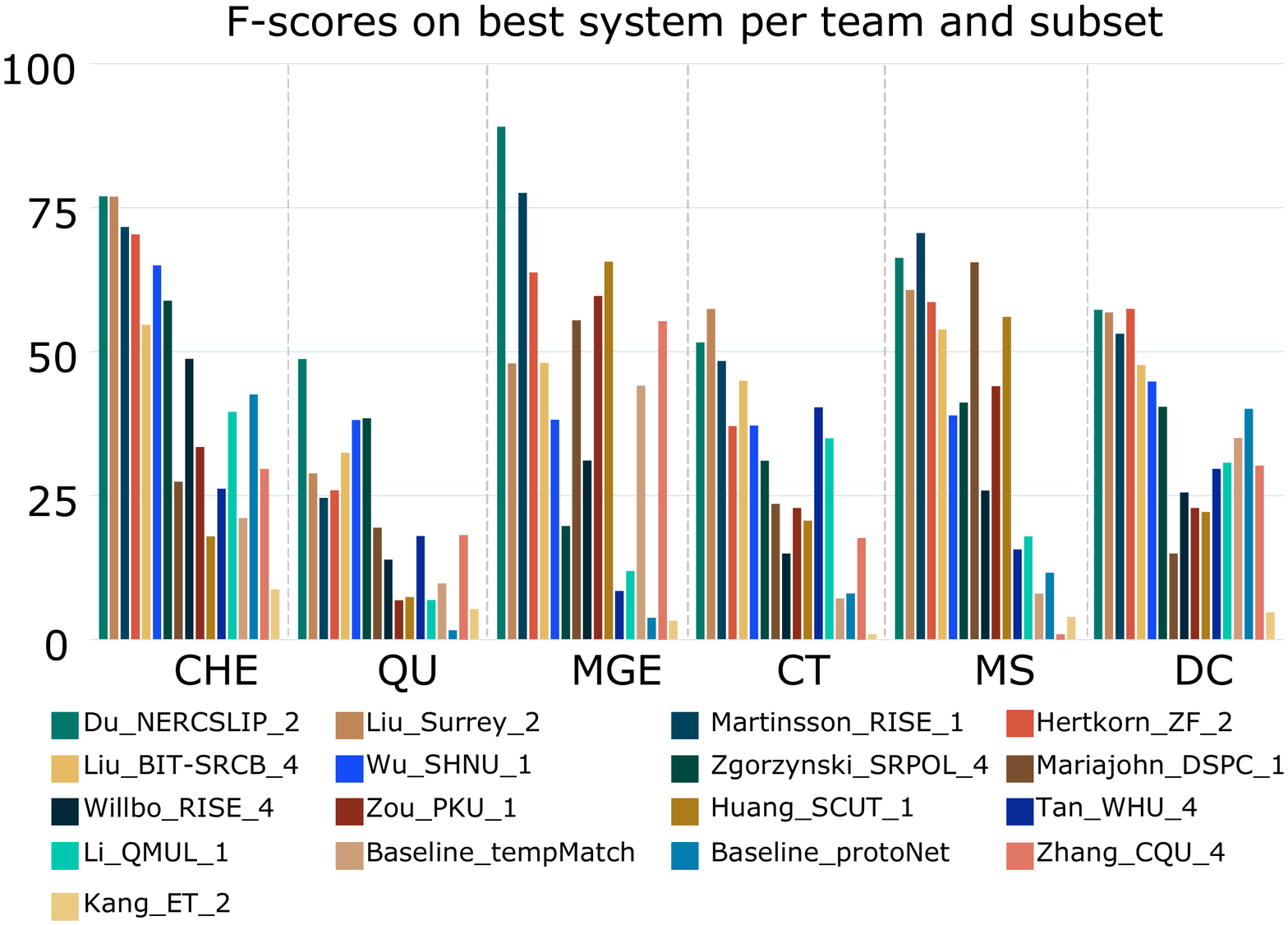}
    \caption{F-Score results by dataset. Systems are ordered by highest scoring rank on the evaluation set.}
    \label{fig:Fscore_dataset}
\end{figure}
\vspace{-0.5cm}
\section{Results}
\label{sec:results}
\vspace{-0.2cm}
For the 2022 edition, 15 teams participated submitting a total of 46 systems.
The results for the highest scoring submission for each team are presented in Table \ref{tab:teams}, together with the reported F-scores on the validation set and summary of the system characteristics. Fig. \ref{fig:Fscore_dataset} presents the F-scores obtained by each team on each subset of the evaluation data.  
The majority of systems adopted a prototypical network approach. Similar to last year's results, simple improvements over the baselines were achieved by applying data augmentation techniques and intelligent post processing. 
Better ways to construct the negative prototype were also explored by some teams who report improved results \cite{Liu2022a, Martinsson2022,  Wu2022,Willbo2022}.
Transductive inference, the method used by the past edition's winning team, was also applied here by several participants \cite{Liu2022a, Li2022, Tan2022,Yang2022}. %Results however do not seem to indicate this option as having an important impact on the results obtained.
The highest scoring system implements a frame-level embedding learning approach which confers to the system a high time resolution capability \cite{Du2022}.  Observing Fig.\ref{fig:Fscore_dataset}, the system was particularly effective on the QU and MGE datasets. This confirms that good time precision is fundamental, particularly for classes with events of very short duration as the ones in these datasets.
The system ranked in second place implements a novel approach designed to optimise the contrast between positive events and negative prototypes \cite{Liu2022a}. This, together with an adaptive segment length dependent on each target class, works well across all the evaluation sets.
The problem of very different length of events across target classes was also directly addressed by other submissions. Both \cite{Martinsson2022} and \cite{Zgorzynski2022} implemented an ensemble approach where each individual model focus on a different input size range.
In \cite{Liu2022} this is explored through a multi-scale ResNet, and in \cite{Willbo2022} with a wide ResNet containing many channels.
Finally, it is worth mentioning the system in \cite{Hertkorn2022}. Their few-shot adaptation was based on fine-tuning alone. The innovation here is related with simple modifications to a CNN based architecture in order to optimise the use of information, particularly in the frequency axis. Furthermore, by allowing the network to overfitt (up to a degree) to the 5 shots, the system achieves surprisingly good performance across all the datasets of the evaluation set.
Overall, this edition saw some novel ideas being implemented that tried to address previously identified challenges related to this task: how to deal with very different event lengths; how to construct a negative class when not explicit labels are given for this; and how to bridge the gap between classification and detection for few shot sound event detection.
We believe these remain relevant questions for our goal and for SED in general, and that the collective work developed here helped pushing few-shot bioacoustic sound event detection into DCASE central stage.

%\section{Discussions}
%\label{sec:disc}
%In this paper, we presented the first few-shot bioacoustic event detection task that organised as part of DCASE 2021 Challenge. We gave details on the datasets used to train and evaluate methods that perform few-shot bioacoustic event detection and also gave details on our proposed baselines, one using spectrogram cross-correlation and another using prototypical networks for sound event detection (SED). Finally, we evaluated the performance of systems submitted to the challenge. We found that best performance can be achieved when using... however, the field of few-shot SED is still novel and there are more advances to be made.

% from the abstract:  insight as to characteristics of the subsets that can influence performance

% 

% -------------------------------------------------------------------------
% Either list references using the bibliography style file IEEEtran.bst
% \clearpage
\bibliographystyle{IEEEtran}
\bibliography{refs}

\end{sloppy}
\end{document}